\documentclass[10pt,twocolumn,letterpaper]{article}

 \usepackage{iccv}              

\usepackage{times}
\usepackage{epsfig}
\usepackage{graphicx}
\usepackage{amsmath}
\usepackage{amssymb}
\usepackage{booktabs}
\usepackage{caption}
\usepackage{color}
\usepackage{multirow}
\usepackage{enumitem}
\usepackage{booktabs}
\usepackage[dvipsnames]{xcolor}
\usepackage[ruled, lined,noend, linesnumbered, commentsnumbered]{algorithm2e}
\usepackage{tikz}

\usepackage{multirow}
\usepackage{colortbl}

\makeatletter
\@namedef{ver@everyshi.sty}{}
\makeatother

\usepackage{amsfonts} 
\usepackage[accsupp]{axessibility}
\DeclareMathSymbol{\shortminus}{\mathbin}{AMSa}{"39}

\newcommand{\PSNR}{PSNR $\uparrow$}

\newcommand{\SSIM}{ SSIM$\uparrow$}
\newcommand{\MRAE}{ SAM$\downarrow$}
\newcommand{\RMSE}{ RMSE$\downarrow$}

\graphicspath{{Figs/}} 

\def\redc{\bf\cellcolor[HTML]{FF999A}}
\def\orangec{\it \cellcolor[HTML]{FFCC99}}
\def\yellowc{\cellcolor[HTML]{FFF8AD}}

\newcommand{\mysection}[1]{\vspace{2pt}\noindent\textbf{#1}}

\definecolor{commentsColor}{RGB}{219, 48, 122}

\SetCommentSty{mycommfont}
\let\oldnl\nl
\newcommand{\nonl}{\renewcommand{\nl}{\let\nl\oldnl}}

\usepackage[normalem]{ulem}
\useunder{\uline}{\ul}{}

\usepackage[precision=2, unit=mm]{lengthconvert}
\usetikzlibrary{patterns}

\definecolor{figblue}{HTML}{1560bd}
\definecolor{figred}{HTML}{a9203e}
\definecolor{figpurple}{HTML}{8844aa}
\definecolor{figgreen}{HTML}{00693e}

\definecolor{iccvblue}{rgb}{0.21,0.49,0.74}
\usepackage[pagebackref,breaklinks,colorlinks,allcolors=iccvblue]{hyperref}

\def\paperID{1269} 
\def\confName{ICCV}
\def\confYear{2025}

\title{UnMix-NeRF: Spectral Unmixing Meets Neural Radiance Fields}

\author{Fabian Perez$^{1,2,\dagger}$, Sara Rojas$^{2}$, Carlos Hinojosa$^{2,*}$, Hoover Rueda-Chac{\'o}n$^{1}$, Bernard Ghanem$^{2}$ \\
$^{1}$Universidad Industrial de Santander \ \ $^{2}$ KAUST
}

\begin{document}

\twocolumn[{%
\renewcommand\twocolumn[1][]{#1}
\maketitle
\begin{center}
\centering
    \vspace{-9pt}
    \includegraphics[width=0.9\linewidth]{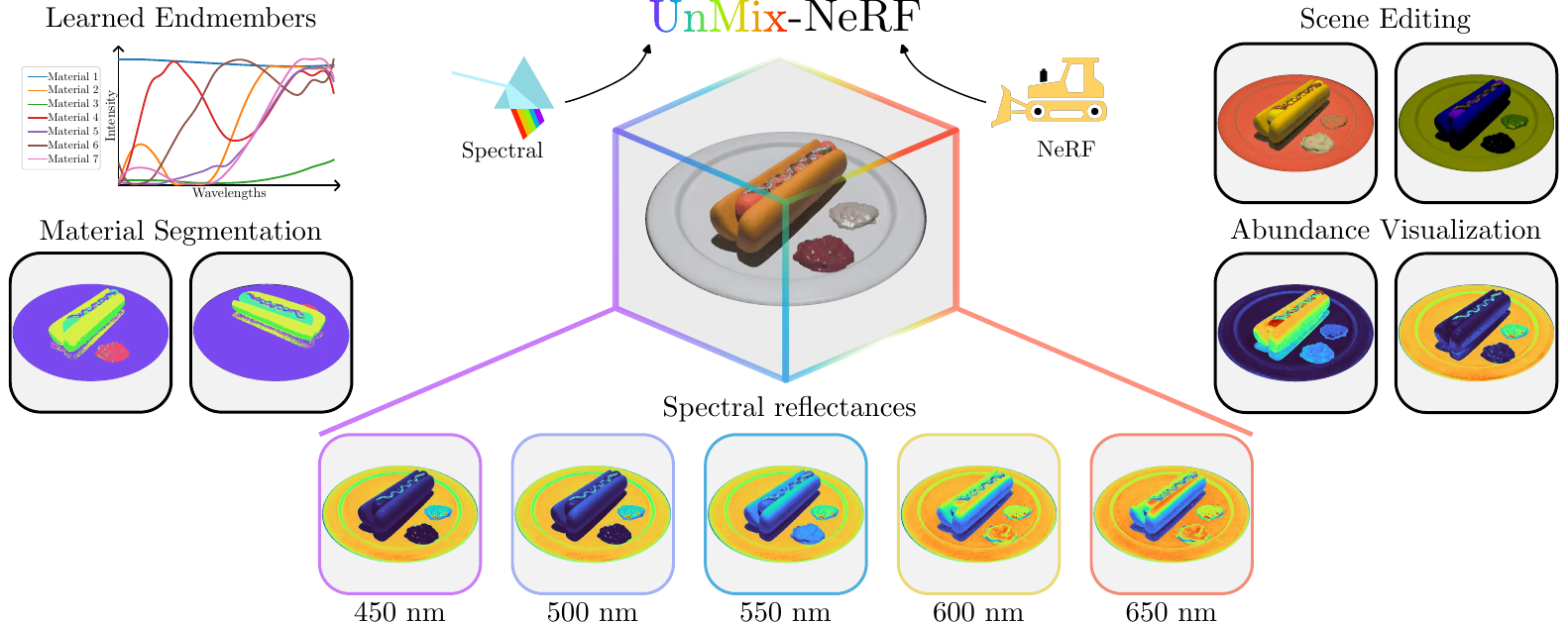}
    \captionof{figure}{UnMix-NeRF: A hyperspectral novel view synthesis framework that leverages spectral unmixing for scene editing and unsupervised material segmentation. By exploiting endmembers and abundances, our approach enables material separation and scene manipulation.}
    \label{fig:intro_fig}
\end{center}%
}]

\def\thefootnote{*}\footnotetext{\vspace{-3pt}Project lead; $\dagger$ Work done during an internship at KAUST.}
\begin{abstract}
Neural Radiance Field (NeRF)-based segmentation methods focus on object semantics and rely solely on RGB data, lacking intrinsic material properties. This limitation restricts accurate material perception, which is crucial for robotics, augmented reality, simulation, and other applications. We introduce UnMix-NeRF, a framework that integrates spectral unmixing into NeRF, enabling joint hyperspectral novel view synthesis and unsupervised material segmentation. Our method models spectral reflectance via diffuse and specular components, where a learned dictionary of global endmembers represents pure material signatures, and per-point abundances capture their distribution. For material segmentation, we use spectral signature predictions along learned endmembers, allowing unsupervised material clustering. Additionally, UnMix-NeRF enables scene editing by modifying learned endmember dictionaries for flexible material-based appearance manipulation. Extensive experiments validate our approach, demonstrating superior spectral reconstruction and material segmentation to existing methods. Project page: \url{https://www.factral.co/UnMix-NeRF}.
\end{abstract}

\section{Introduction}
\label{sec:intro}


 

%

Neural Radiance Fields (NeRF) \cite{mildenhall2021nerf} have revolutionized 3D scene representation, enabling photorealistic novel view synthesis through implicit volumetric rendering. Recent advances have extended NeRF to other tasks such as segmentation~\cite{engelmann2024opennerf}, object detection~\cite{hu2023nerf}, and spatial reasoning~\cite{kerr2023lerf}, leveraging pre-trained vision models such as SAM \cite{kirillov2023segment} and CLIP \cite{radford2021learning} for higher-level scene understanding. Among these, semantic segmentation has emerged as a powerful tool, enabling the grouping of regions into object-level categories in multi-view settings \cite{liu2024sanerf}. A more challenging task lies in understanding and segmenting materials, moving beyond recognizing what an object is to understanding what it is composed of. Materials are fundamental to how we interpret and interact with the world, as they define surface properties, functionality, and usability, enabling richer scene understanding and making them essential for applications such as robot autonomy \cite{adarsh2016performance}, augmented reality \cite{zheng2024materobot}, and simulation \cite{vecchio2024matfuse}, where accurate material perception is crucial for decision-making and interaction.

Most existing 3D segmentation approaches focus on semantic segmentation and rely solely on RGB data, which captures color and geometric structure but lacks intrinsic material properties. This limitation can lead to issues like metamerism, where two objects appear identical under a particular light source despite having different materials and spectral radiance distributions. Furthermore, many materials exhibit distinct spectral behaviors outside the visible spectral range, such as in the near-infrared, where we can find indicators for vegetation health \cite{kureel2022modelling}, or in the ultraviolet, where cues about minerals can be extracted from fluorescence \cite{lakowicz2006principles}; information that RGB data fails to capture.



Spectral imaging (SI) emerges as a powerful framework for scene representation, as it provides rich spectral information across multiple wavelengths \cite{huang2022spectral,hinojosa2021fast,correa2017multiple}.  This, in turn, allows for a more complete characterization of a scene, effectively encoding unique spectral cues for different materials. Integrating spectral data into implicit neural representations remains a challenge. Standard NeRF formulations model scene appearance using only RGB values, which inherently limits their capacity to differentiate materials with similar color but distinct spectral properties. While recent efforts have attempted to extend NeRF to spectral data \cite{li2024spectralnerf, chen2024hyperspectral} by replacing the standard RGB output with per-channel spectral radiance predictions, these approaches treat spectral information as an additional dimension in the output space without explicitly leveraging its inherent structure, e.g. materials have representative spectral signatures.


In real-world scenes, the spectral signature at a given point often arises from a mixture of multiple materials rather than a single homogeneous substance. Spectral unmixing \cite{keshava2002spectral} is the process of decomposing these mixed signals into their constituent materials, represented as a set of pure spectral signatures (\textit{endmembers}), and estimating their corresponding \textit{abundances}, which define the proportion of each endmember at a given 3D point. Consequently, spectral unmixing serves as a fundamental tool for distinguishing materials within a scene \cite{wetherley2017mapping}. However, to the best of our knowledge, there is currently no work in view synthesis that fully exploits the rich structure of spectral data, not only to enhance the quality of novel view generation but also to enable more precise and physically meaningful identification of materials within a scene.

\mysection{Contributions.} In this paper, we propose UnMix-NeRF, the first framework that integrates spectral unmixing into NeRF, allowing joint spectral novel view synthesis and unsupervised material segmentation. Our method learns global endmembers through a dictionary optimized during training and models spectral reflectance via a diffuse and specular decomposition. The diffuse component is represented by per-point abundances and scaling factors, while a dedicated branch predicts view-dependent specular effects. The final spectral radiance is obtained by combining both components. Moreover, the per-point spectral unmixing naturally enables unsupervised material segmentation by leveraging the rendered abundance vectors. To summarize, our key contributions are:
\begin{itemize}
    \item[(i)] We introduce UnMix-NeRF, a framework for hyperspectral novel view synthesis that leverages spectral unmixing to achieve high-quality spatial and spectral renderings by encoding material properties from a multi-view 3D hyperspectral dataset.
    \item[(ii)] Our framework inherently enables unsupervised segmentation by leveraging the rendered abundance vectors learned during training, which accurately delineate spatially varying material properties within scenes.
    \item[(iii)] Exploiting the learned endmember dictionary, our framework supports scene editing through direct and customizable manipulation of scene appearance with fine-grained control over individual materials.
\end{itemize}

Our approach not only outperforms state-of-the-art methods in hyperspectral view synthesis but also achieves pixel-accurate 3D material segmentation. Also, we extend a synthetic hyperspectral dataset with paired 3D material annotations, enabling rigorous evaluation of spectral unmixing and segmentation in complex scenes.

\vspace{-0.05in}
\section{Related Work}
\label{sec:relatedwork}
\mysection{Neural Radiance Fields.}
NeRF revolutionized novel view synthesis by representing scenes as continuous neural fields. Through volumetric rendering and multi-view optimization, it learns to predict density and color at any 3D point and viewing direction, enabling photorealistic view synthesis \cite{mildenhall2021nerf}. Several works have extended NeRF to new imaging modalities: SeaThru-NeRF \cite{levy2023seathru} models underwater light scattering, Thermal-NeRF \cite{ye2024thermal} handles infrared emissions, LiDAR-NeRF \cite{tao2024lidar} enables LiDAR point cloud synthesis, and NeSpoF \cite{kim2023neural} incorporates polarization information. For hyperspectral imaging, HS-NeRF \cite{chen2024hyperspectral} and Spec-NeRF \cite{li2024spec} directly extend NeRF to predict per-wavelength radiance. Recent approaches like SpectralNeRF \cite{li2024spectralnerf}, introduce physically-based spectral rendering by extending NeRF with an additional autoencoder to predict per-wavelength radiance, and HyperGS \cite{thirgood2024hypergs}, a concurrent work that adapts 3D Gaussian Splatting for spectral scene reconstruction. However, these methods treat spectral bands as independent channels, failing to take advantage of the rich material information encoded within the spectral signatures. These extensions require substantial computational overhead and also ignore the structured nature of spectral information, limiting their ability to disentangle materials.

\mysection{Spectral Unmixing.}
Spectral unmixing decomposes pixels into endmembers (materials signatures) and their abundances (percentage of material per pixel) \cite{keshava2002spectral}. Beyond its use in remote sensing, unmixing enables material understanding in indoor and outdoor scenes in robotics \cite{dieters2024robot}, industrial inspection \cite{gat1997spectral}, and material classification \cite{zhao2019laboratory}. The process typically involves identifying the different concurrent endmember spectra and their fractional abundances, per pixel, while enforcing physical constraints like non-negativity and sum-to-one (i.e., the sum of all abundances must be 100\%). Classical approaches employ the linear mixing model (LMM), which assumes a linear combination of endmembers \cite{keshava2002spectral}. However, LMM struggles with spectral variability caused by illumination, atmospheric conditions, and material mixing effects \cite{borsoi2021spectral}. Extensions like the Augmented LMM (ALMM) \cite{hong2018augmented} and the Extended LMM (ELMM)~\cite{veganzones2014new} address this by incorporating scaling factors and environmental variations through learned dictionaries and bundle-based representations, improving abundance estimation accuracy under real-world conditions. Recent deep learning approaches leverage transformers and convolutional neural networks (CNNs) to learn endmembers and abundances directly from data \cite{ghosh2022hyperspectral, zhang2018hyperspectral}. While these methods have shown promise for 2D spectral unmixing, they are currently limited to single-view analysis and have not been extended to 3D reconstruction, where material decomposition could enable richer scene understanding. The challenge of incorporating physical unmixing constraints into neural architectures for 3D scenes remains unexplored.

\mysection{Material Segmentation.} 
Recent advances in self-supervised learning and foundation models have revolutionized unsupervised segmentation. Methods like DINO \cite{oquab2023dinov2} and SAM \cite{kirillov2023segment} learn powerful visual features that allow zero-shot segmentation without task-specific training. These approaches leverage large-scale pretraining to discover semantic boundaries and object-centric representations, reducing the need for manual annotations. Materialistic \cite{sharma2023materialistic} builds on this foundation, enabling interactive material selection based on a user-provided pixel, showing robustness to shading and geometric variations.
For NeRF, several works explore unsupervised object segmentation through multi-view consistency. RFP \cite{liu2022unsupervised} and ONeRF \cite{liang2022onerf} segment objects by propagating visual features through the radiance field using photometric consistency. More recent approaches like LERF \cite{kerr2023lerf} and OpenNeRF \cite{engelmann2024opennerf} leverage language-vision models to enable open-vocabulary 3D segmentation. However, these methods focus on semantic segmentation rather than material understanding, as they rely on RGB appearance features that can be ambiguous for material discrimination. Recent works like MaterialSeg3D \cite{li2024materialseg3d} attempt material segmentation in 3D but still operate on RGB images using semantic priors learned from labeled datasets.
Concurrent work SAMa \cite{fischer2024sama} requires a user-provided click and is limited to RGB data. Hyperspectral imaging, with its rich spectral information, naturally enables unsupervised material segmentation due to the inherent discriminative power of spectral signatures \cite{perez2024beyond}, thus eliminating the ambiguity associated with RGB features.

\section{Preliminary}

In this section, we review the NeRF volumetric rendering formulation and provide an overview of spectral unmixing.

\begin{figure*}[t!]
    \centering
    \centerline{\includegraphics[width=\linewidth]{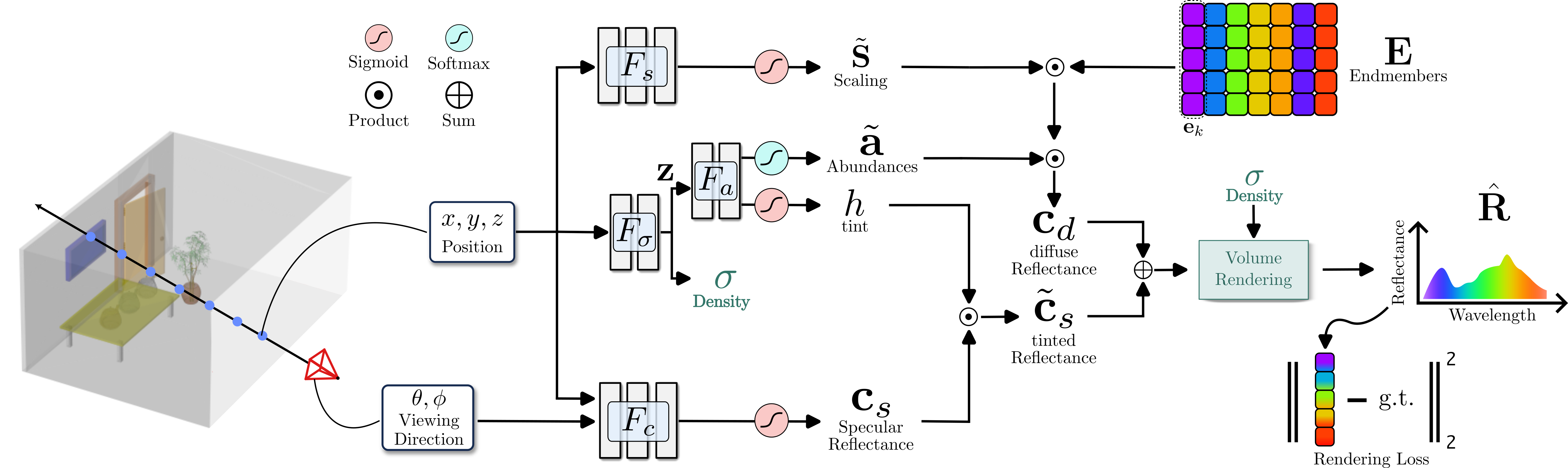}}
    \caption{\textbf{Overview of NeRF unmixing approach.}
    Given a viewing direction $(\theta, \phi)$ and 3D coordinates $(x,y,z)$, our network predicts a density $\sigma$ and a set of abundance vectors $\tilde{\mathbf{a}}$, along with scalar factors $\tilde{\mathbf{s}}$ that scale the spectral endmembers $\mathbf{E}$. The scaled endmembers, when multiplied by the predicted abundances $\tilde{\mathbf{a}}$, yield the diffuse reflectance component $\mathbf{c}_d$. 
    In parallel, a specular reflectance $\mathbf{c}_s$ and an additional tint factor $h$ are also predicted to capture local illumination effects. 
    The final spectral radiance is obtained by combining these diffuse and specular components $\mathbf{c}_{d} + \tilde{\mathbf{c}}_{s}$ under volumetric rendering with the predicted density. 
    An $\ell_2$ loss on the reconstructed spectral signatures is sufficient to guide unmixing and specular-diffuse modeling. 
   } 
    \label{fig:method}
\end{figure*}

\subsection{Neural Radiance Fields}
\label{sec:nerf}
In NeRF, a scene is represented as a continuous volumetric function $F$ parameterized by a Multi-Layer Perceptron (MLP). For any 3D point $\mathbf{x} = (x,y,z)$ and viewing direction $\mathbf{d} = (\theta, \phi)$, the network predicts a density $\sigma$ and a view-dependent RGB color $\mathbf{c}$ as $(\sigma, \mathbf{c})=F(\mathbf{x}, \mathbf{d})$.
To render an image from a novel viewpoint, NeRF performs volumetric rendering by sampling points along each camera ray, defined as  $\mathbf{r}(t) = \mathbf{o} + t\mathbf{d}$, where $\mathbf{o} \in \mathbb{R}^{3}$ is the ray origin and $\mathbf{d} \in \mathbb{R}^{3}$ is the unit direction vector. The radiance is then accumulated using the classical volume rendering equation \cite{mildenhall2021nerf}.
%
In practice, this integral is approximated as 
\begin{equation}
   C(\mathbf{r}) = \sum_{i=1}^{N_t} T_i(1-\exp(-\sigma_i\delta_i))\mathbf{c}_i,
   \label{eq:vol_rendering}
\end{equation}
where $T_i = \exp(-\sum_{j=1}^{i-1}\sigma_j\delta_j)$, $\delta_i = t_{i+1} - t_i$ is the distance between adjacent samples, and $N_t$ is the total number of samples. The network is optimized by minimizing the mean squared error between rendered and ground truth pixel colors across multiple training views:
\begin{equation}
   \mathcal{L} = \sum_{\mathbf{r} \in \mathcal{R}} \|C(\mathbf{r}) - C^*(\mathbf{r})\|_2^2 ,
\end{equation}
where $\mathcal{R}$ represents the set of camera rays from all training views and $C^*$ denotes the corresponding ground truth.


\subsection{Spectral Unmixing}
\label{sec:unmixing}
A hyperspectral image with $B$ spectral bands and $H\times W$ spatial pixels can be represented as $\mathbf{Y} = [\mathbf{y}_1,\cdots,\mathbf{y}_n,\cdots,\mathbf{y}_N] \in \mathbb{R}^{B\times N}$, where $\mathbf{y}_n \in \mathbb{R}^{B\times 1}$ is a spectral pixel, for $n=1, \ldots, N$, with $N=HW$ as the total number of pixels. Let $\mathbf{E} = [\mathbf{e}_1,\cdots,\mathbf{e}_k, \cdots, \mathbf{e}_K] \in \mathbb{R}^{B\times K}$ be the endmember matrix, containing the spectral signatures of $K$ pure materials. The Linear Mixing Model (LMM) \cite{keshava2002spectral} assumes that each spectral pixel is a linear combination of these endmembers,
\begin{equation}
   \mathbf{y}_n = \mathbf{E}\mathbf{a}_n + \boldsymbol{\epsilon}_n,
\end{equation}
where $\mathbf{a}_n \in \mathbb{R}^K$ is the abundance vector associated with pixel $n$-th, quantifying the contribution of each endmenmber and $\boldsymbol{\epsilon}_n$ is the residual (error approximation) term. The abundance coefficients satisfy the following constraints:
\begin{equation}\label{eq:4}
   \mathbf{a}_n \succeq 0, \quad \mathbf{1}^T\mathbf{a}_n = 1,
\end{equation}
ensuring that abundances are non-negative and that they sum to one, enforcing a physically meaningful representation of spectral mixing.

To account for spectral variability caused by the environment and illumination, the Extended Linear Mixing Model (ELMM) \cite{veganzones2014new} introduces pixel-wise scaling factors $\mathbf{S}_n \in \mathbb{R}^{K\times K}$,
\begin{equation}
\mathbf{y}_n = \mathbf{E} \mathbf{S}_n \mathbf{a}_n + \boldsymbol{\epsilon}_n,
\end{equation}
where $\mathbf{S}_n$ is a diagonal matrix with non-negative entries.
\section{Proposed Method}
\label{sec:method}





Figure \ref{fig:method} illustrates our framework for hyperspectral novel view synthesis and material segmentation, which integrates spectral unmixing with volumetric rendering. In our approach, given a viewing direction $(\theta, \phi)$ and a 3D point $(x,y,z)$, the network predicts a density $\sigma$ together with a set of abundance vectors and scalar factors that scale a set of spectral endmembers. These scaled endmembers, when combined with the predicted abundances, form the diffuse reflectance component $\mathbf{c}_d$. In parallel, the model estimates a specular reflectance $\mathbf{c}_s$ and an additional tint factor $h$ to account for local illumination effects. By blending these diffuse and specular components through volumetric rendering, the framework generates the final spectral radiance, which can be mapped from the scene coordinate space to sRGB for image synthesis (see, e.g., \cite{verbin2022ref, hedman2021baking}). Moreover, the per-point spectral unmixing naturally enables unsupervised material segmentation by leveraging the rendered abundance vectors.



\subsection{Spectral Unmixing Field}
\label{subsec:unmixing_field}

We extend the standard NeRF formulation by incorporating the ELMM at each 3D point (see section \ref{sec:unmixing}). Let $\mathbf{x}=(x,y,z)$ be a coordinate in the scene. We first predict a density value along with an intermediate latent feature via an MLP, denoted as $(\sigma, \mathbf{z})=F_{\sigma}(\mathbf{x})$, which maps $\mathbf{x}$ to a scalar density $\sigma \ge 0$ and a latent feature vector $\mathbf{z}\in\mathbb{R}^{D}$, where $D$ is the dimensionality of the feature vector. This density is used for volumetric rendering (Section~\ref{subsec:volume_rendering}), while $\mathbf{z}$ serves as input for subsequent predictions.

Next, we introduce a learnable dictionary of $K$ endmembers $\mathbf{E} \in \mathbb{R}^{B \times K}$, where each column $\mathbf{e}_k\in\mathbb{R}^{B}$ represents the pure spectral signature of the $k$-th material. These global endmembers are jointly optimized with the network parameters and can be initialized either by standard 2D unmixing methods (e.g., using vertex component analysis (VCA) \cite{nascimento2005vertex}) or randomly.

To model local spectral mixing, we predict two components at each point: (i) a set of $K$ \emph{scaling factors} to account for illumination and environmental variability, and (ii) a set of $K$ \emph{abundances} representing the fractional presence of each material. Specifically, for (i), a dedicated MLP is used to obtain the scaling factors as $\mathbf{s} = F_{s}(\mathbf{x})$, with $\mathbf{s}\in \mathbb{R}^K$. To ensure non-negativity, we apply a sigmoid function to $\mathbf{s}$, hence obtaining $\tilde{\mathbf{s}} = \operatorname{sigmoid}(\mathbf{s})$, with each element $\tilde{s}_k \in [0,1]$. (ii) The intermediate feature $\mathbf{z}$ obtained from $F_{\sigma}(\cdot)$ is then used by an abundance head $F_{a}(\cdot)$ to predict raw abundances and a tint factor (see Section \ref{subsec:specular_field}), expressed as $(\mathbf{a}, h)=F_{a}(\mathbf{x})$, where $\mathbf{a} \in \mathbb{R}^K$ and $h \in \mathbb{R}$. A softmax activation with temperature $\tau>0$ is applied to $\mathbf{a}$, and computed as follows:
\begin{equation}
\tilde{a}_k = \frac{\exp\!\left(a_k/\tau\right)}{\sum_{j=1}^{K} \exp\!\left(a_j/\tau\right)},
\end{equation}
thus ensuring non-negativity and the sum-to-one constraint according to Eq. \eqref{eq:4}. Additionally, $h$ is passed through a sigmoid activation to constrain its values within $[0,1]$, thus modulating the contribution of the diffuse color. Each abundance $\tilde{a}_k$ corresponds to the fractional presence of the $k$-th material at point $\mathbf{x}$, providing a physically meaningful decomposition of the current spectral signature. With these, the diffuse reflectance is computed as follows:
\begin{equation}
\mathbf{c}_{d} = \mathbf{E} \, \tilde{\mathbf{S}} \tilde{\mathbf{a}},
\end{equation}
where $\tilde{\mathbf{S}}$ is a diagonal matrix with non-negative scaling factors $\tilde{\mathbf{s}}$ on its diagonal.


\subsection{Specular Field}
\label{subsec:specular_field}

Since spectral unmixing is inherently designed for diffuse reflectance, it fails to capture view-dependent effects such as specular highlights. To address this limitation, we introduce a specular field. Motivated by Ref-NeRF \cite{verbin2022ref}, we leverage the tint factor $h$ predicted by the abundance head $F_a(\cdot)$ (see Section \ref{subsec:unmixing_field}) to modulate the specular reflectance, allowing for better control of view-dependent effects. Then, taking both $\mathbf{x}$ and $\mathbf{d}$ as input, an MLP $F_c(\cdot)$ is used to obtain the specular reflectance vector as $\mathbf{c}_{s} = F_c(\mathbf{x},\mathbf{d})$, where $\mathbf{c}_{s} \in \mathbb{R}^{B}$. The tint factor is then used to modulate this specular term as
\begin{equation}
\tilde{\mathbf{c}}_s=h\mathbf{c}_s.
\end{equation}
The final per-point spectral reflectance is obtained by combining the diffuse and specular components under the dichromatic model \cite{cook1982reflectance}:
\begin{equation}
\mathbf{c} = \mathbf{c}_{d} + \tilde{\mathbf{c}}_{s}.    
\end{equation}

\subsection{Volume Rendering and Camera Response}
\label{subsec:volume_rendering}

Following the standard NeRF formulation (see Section \ref{sec:nerf}), we sample points along camera rays and accumulate the spectral radiance. The accumulated spectral radiance along a ray is computed using Eq. \eqref{eq:vol_rendering}. Similarly, we can volume-render the material abundances to obtain the per-ray abundance vectors as:
\begin{equation}
A(\mathbf{r}) = \sum_{i=1}^{N_t} T_i \, \bigl(1 - \exp({-\sigma_i \delta_i})\bigr)\, \tilde{\mathbf{a}}_i,
\end{equation}
for the $i$-th sample location $\mathbf{x}_i$ along the ray $\mathbf{r}$. This allows us to render the abundances for each point in the scene



To recover an RGB image from the rendered hyperspectral signature, we employ a spectral response model. This conversion involves transforming the spectral power distribution in the scene coordinate space to a device-independent color space (such as XYZ) and then to the target sRGB color space. Let $\mathbf{M}\in\mathbb{R}^{3 \times B}$ be the camera spectral response matrix that encapsulates both the color matching functions and the display primaries; then, the RGB color is given by
\begin{equation}
    C_{\mathrm{rgb}}(\mathbf{r}) = \mathbf{M}\,C(\mathbf{r}).
\end{equation}
This approach ensures accurate color reproduction by modeling the entire imaging pipeline from scene radiance to displayed color.

\subsection{Material Segmentation via Cluster Probe}

For unsupervised material segmentation, we perform a cluster assignment for each ray's predicted spectral signature by employing the learned spectral endmembers $\mathbf{E}$ as cluster centers. Given the predicted spectral signature for each ray $C(\mathbf{r}) \in \mathbb{R}^{B}$, we compute the normalized inner product against the dictionary of endmembers as
\begin{equation} 
    \mathbf{p}(\mathbf{r}) = \text{softmax}\left(\frac{\mathbf{E}^\top C(\mathbf{r})}{|\mathbf{E}||C(\mathbf{r})|}\right),
\end{equation}
where $\mathbf{p}(\mathbf{r}) \in \mathbb{R}^{K}$ contains the probabilities indicating cluster memberships for each of the $K$ materials. Finally, material segmentation is obtained by assigning each ray $\mathbf{r}$ to the material cluster with the highest probability following
\begin{equation} 
    m(\mathbf{r}) = \arg\max_{k} p_{k}(\mathbf{r}). 
\end{equation}

\subsection{Loss Function}
\label{subsec:loss}

We jointly optimize all parameters, including the endmember matrix $\mathbf{E}$, and the weights of the MLPs for density ($F_\sigma$), scaling ($F_s$), abundance ($F_a$), and specular reflectance ($F_c$), by minimizing a combination of hyperspectral and RGB reconstruction errors. Let $C(\mathbf{r})$ and $C_{\mathrm{rgb}}(\mathbf{r})$ denote the rendered hyperspectral and RGB predictions for ray $\mathbf{r}$, and let $C^{*}(\mathbf{r})$ and $C^{*}_{\mathrm{rgb}}(\mathbf{r})$ be the corresponding ground-truth measurements.
The losses are defined as
\begin{align}
    L_{\mathrm{spec}} &= \sum_{\mathbf{r}\in\mathcal{R}} \left\|C(\mathbf{r}) - C^{*}(\mathbf{r})\right\|_2^2,\\
    L_{\mathrm{rgb}} &= \sum_{\mathbf{r}\in\mathcal{R}} \left\|C_{\mathrm{rgb}}(\mathbf{r}) - C^{*}_{\mathrm{rgb}}(\mathbf{r})\right\|_2^2,
\end{align}
where $\mathcal{R}$ is the set of all training rays. Then, our final objective is given by
\begin{equation}
    L = \lambda_{\mathrm{spec}}\,L_{\mathrm{spec}} + \lambda_{\mathrm{rgb}}\,L_{\mathrm{rgb}},
\end{equation}
with hyperparameters $\lambda_{\mathrm{spec}}$ and $\lambda_{\mathrm{rgb}}$ balancing the importance of hyperspectral versus RGB fidelity. Minimizing this loss not only leads to accurate novel view synthesis but also ensures physically consistent spectral unmixing. This results in both high-quality spectral reconstructions and reliable material segmentation through the per-ray abundance predictions.

\subsection{Implementation Details}
\label{subsec:implementation}

We implemented our method within the Nerfstudio framework \cite{tancik2023nerfstudio}, building on top of the Nerfacto implementation. The hidden feature dimension $D$ was set to 16, and we integrated Nerfacc \cite{li2023nerfacc} into the volumetric rendering and sampling process to reduce training time. Additionally, we applied gradient scaling based on the squared ray distance to each pixel, as suggested in \cite{lin2023gradientscaling}. For optimization, we used the Adam optimizer with a learning rate of $1\times10^{-2}$ and an exponential learning rate scheduler. All models were trained for $20,000$ iterations, and all experiments were performed on an NVIDIA A100 GPU, with a maximum memory usage of approximately 20 GB. For all experiments, we set \(\lambda_{\mathrm{spec}} = 5\) and \(\lambda_{\mathrm{rgb}} = 1\). To ensure physically meaningful spectral signatures, the endmembers \(  \mathbf{E} \) were constrained at each iteration by applying element-wise clamping within the range \([0,1]\). For each trained scene, the number of endmembers \( K \) is selected as a heuristic approximation of the number of distinct materials present in the scene.

\section{Experimental Results}
\label{sec:results}

We extensively evaluate the performance of our proposed approach using standard metrics and compare it with state-of-the-art methods.

\subsection{Datasets, Preprocessing, Metrics and Baselines} 
\mysection{Datasets.} Three different datasets are employed for hyperspectral novel view synthesis and material segmentation: \begin{itemize} 
\item \textbf{NeSpoF Dataset~\cite{kim2023neural}.} We use the NeSpoF spectro-polarimetric dataset, focusing exclusively on the $S_0$ Stokes component representing spectral radiance. Specifically, we utilize the synthetic dataset comprising $21$ spectral bands ranging from $450$ nm to $650$ nm. This dataset includes four distinct scenes: \textit{ajar}, \textit{hotdog}, \textit{cbox\_dragon}, and \textit{cbox\_sphere}. We adopt the provided camera poses and their calibrated RGB projections.  Additionally, we extended the dataset by generating material maps from the synthetic scenes, obtaining ground-truth labels for material segmentation. Further details about the extension are provided in the supplementary material.

\item \textbf{Surface Optics Dataset~ \cite{chen2024hyperspectral}.} The Surface Optics dataset contains real-world imagery captured across $128$ spectral bands, spanning wavelengths from $370$ nm to $1100$ nm. This dataset consists of four object-centric scenes: Rosemary, Basil, Tools and Origami. We adopt the original camera poses.

\item \textbf{BaySpec Dataset~\cite{chen2024hyperspectral}.} The BaySpec dataset offers hyperspectral data across $140$ spectral bands, covering a wavelength range from $400$ nm to $1110$ nm. BaySpec offers three scenes: Pinecone, Caladium and Anacampseros and includes measured camera poses as described in the original dataset publication.
\end{itemize}


\mysection{Metrics.} We evaluate UnMix-NeRF and competing methods on hyperspectral view synthesis using rigorous metrics that capture both accuracy and spectral fidelity, including PSNR, SSIM, Spectral Angle Mapping (SAM)~\cite{kruse1993spectral}, and RMSE. For unsupervised material segmentation, we report standard metrics, including mean Intersection over Union (mIoU) and F1 score.


\begin{figure*}[t!]
    \centering
    \centerline{\includegraphics[width=\linewidth]{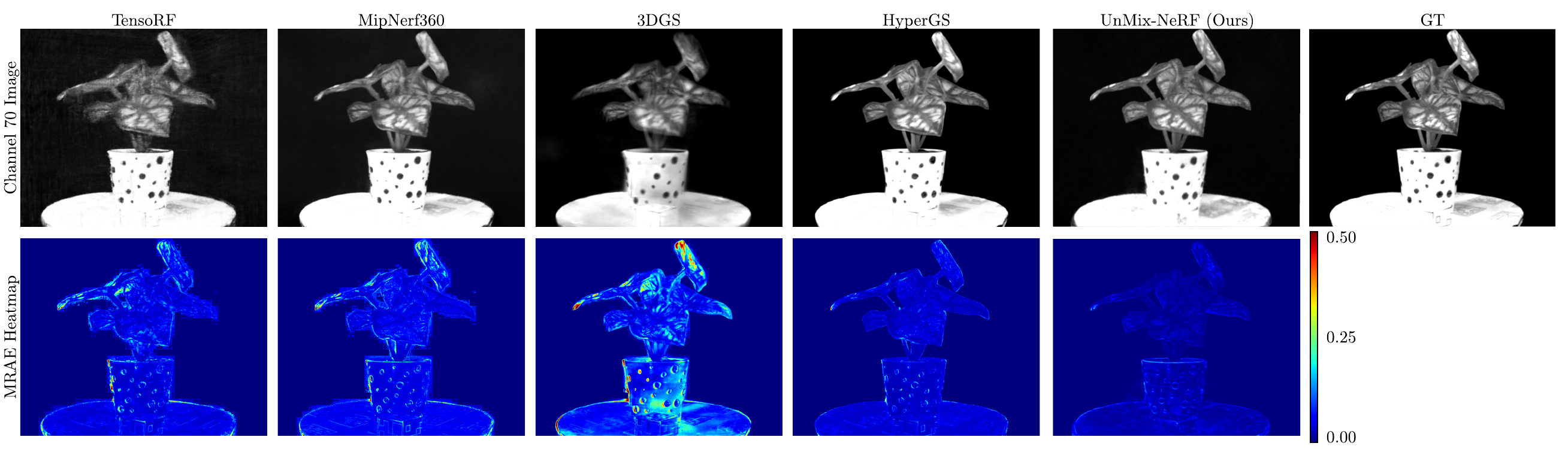}}
    \caption{Visualization of the top-4 performing methods for frame 51 out of 359 for the Caladium plant scene from the Bayspec dataset. The top row shows the 70th spectral channel out of the 141-channel predicted image, and the bottom row provides a raw pixel-wise mean relative absolute error heatmap of the scene.}
    \label{fig:errors}
\end{figure*}

\mysection{Baselines.} We compare our method against existing novel view methods available for hyperspectral imaging. Specifically, we compare with NeSpoF \cite{kim2023neural}, originally designed for spectro-polarimetric imaging, restricting our comparison to the spectral component only ($S_0$ Stokes). Also, we include HS-NeRF, which extends the NeRF framework from the conventional RGB 3-channel output to a multi-channel spectral representation. Lastly, we also compare against HyperGS, a framework for hyperspectral novel view synthesis based on the 3D Gaussian Splatting (3DGS). Also, we include comparisons with traditional NeRF models adapted for hyperspectral data (NeRF, MipNeRF, TensoRF, Nerfacto and MipNeRF-360) for the Bayspec and Surface Optics datasets.

\subsection{Ablation Studies}
To evaluate the effectiveness and impact of individual components in our UnMix-NeRF model, we performed extensive ablation studies. Table 1 summarizes the results obtained from each ablation step performed over the NeSpoF dataset. The initial baseline model (Unmix Only) predicts abundances using global endmembers without further constraints or enhancements. We then sequentially added:
\begin{enumerate}
    \item Physical constraints, from Section 3.1, using activation functions to enforce realistic abundance estimates.
    \item Scaling factors, in line with the ELMM, enable the model to better represent local spectral variability.
    \item An RGB loss term consisting of a spectral-to-RGB projection that helps with realism and spectral consistency.
    \item The VCA initialization of the global endmembers instead of random initialization.
    \item Finally, the specular field component is used to accurately model specular reflectances via a dichromatic model.
\end{enumerate}

\begin{table}[t]
    \centering
      \resizebox{\columnwidth}{!}{
    \begin{tabular}{lcccc}
        \toprule
        \textbf{Ablation Step} & \textbf{PSNR}~$\uparrow$ & \textbf{SSIM}~$\uparrow$ & \textbf{SAM}~$\downarrow$ & \textbf{RMSE}~$\downarrow$ \\
        \midrule
        Base: Unmix Only      & 24.73 & 0.710 & 0.063 & 0.058\\
        + Physical Constraint & 22.36 & 0.722 & 0.075 & 0.076 \\
        + Scaling Factors     &29.36 & 0.914 & 0.026 & 0.034 \\
        + RGB Loss            & 30.92 & 0.923 & 0.025 & 0.028 \\
        + VCA Initialization   & 31.09 & 0.923 & 0.024 & 0.027 \\
        + Specular Field      & $\mathbf{33.20}$ & $\mathbf{0.935}$ & $\mathbf{0.022}$ & $\mathbf{0.023}$ \\
        \bottomrule
    \end{tabular}
          }
    \caption{Ablation study on the NeSpoF dataset, evaluating the impact of each component in UnMix-NeRF. Metrics include PSNR ($\uparrow$), SSIM ($\uparrow$), Spectral Angle Mapper (SAM, $\downarrow$), and RMSE ($\downarrow$). Each row sequentially adds a component to the baseline model. The final configuration, incorporating all elements, achieves the best overall performance.}
    \label{tab:ablation}
\end{table}

\noindent Results in Table \ref{tab:ablation} illustrate the incremental contributions of each component added to our UnMix-NeRF baseline. The initial model (Unmix Only) achieves moderate spectral reconstruction quality (PSNR $=24.73$), establishing a performance baseline without additional constraints. Introducing physical constraints reduces the PSNR slightly by a margin of $-2.37$ due to the restrictive nature of these constraints on spectral variability. The addition of scaling factors significantly boosts the performance (PSNR increases by $+7.00$, SAM decreases by $-0.049$, and RMSE decreases by $-0.042$), underscoring their importance in modeling local spectral variations. Incorporating the RGB loss further improves the spectral accuracy (PSNR improves by $+1.56$), demonstrating the benefit of coupling hyperspectral predictions with RGB supervision. Using the VCA initialization provides a minor improvement of $+0.17$ in PSNR, indicating that an informed initialization of the endmembers aids the optimization stability. Finally, integrating the specular field yields the best overall results (PSNR increases by $+2.11$, SSIM reaches $0.935$), confirming that explicitly modeling specular reflectance substantially enhances spectral rendering and overall image quality.

\subsection{Qualitative Results}

To further assess the effectiveness of our UnMix-NeRF framework, we provide qualitative comparisons against state-of-the-art methods in hyperspectral novel view synthesis. Figure~\ref{fig:errors} illustrates the quality of reconstruction between different methods in the Caladium scene of the Bayspec dataset. The top row displays the 70th spectral channel from the reconstructed hyperspectral cube, while the bottom row depicts the corresponding mean relative absolute error (MRAE) heatmaps, highlighting pixel-wise reconstruction errors. Our method achieves the most accurate spectral predictions, significantly reducing reconstruction artifacts and preserving fine-grained spectral details. 

\begin{figure}[t]
    \centering
    \includegraphics[width=\linewidth]{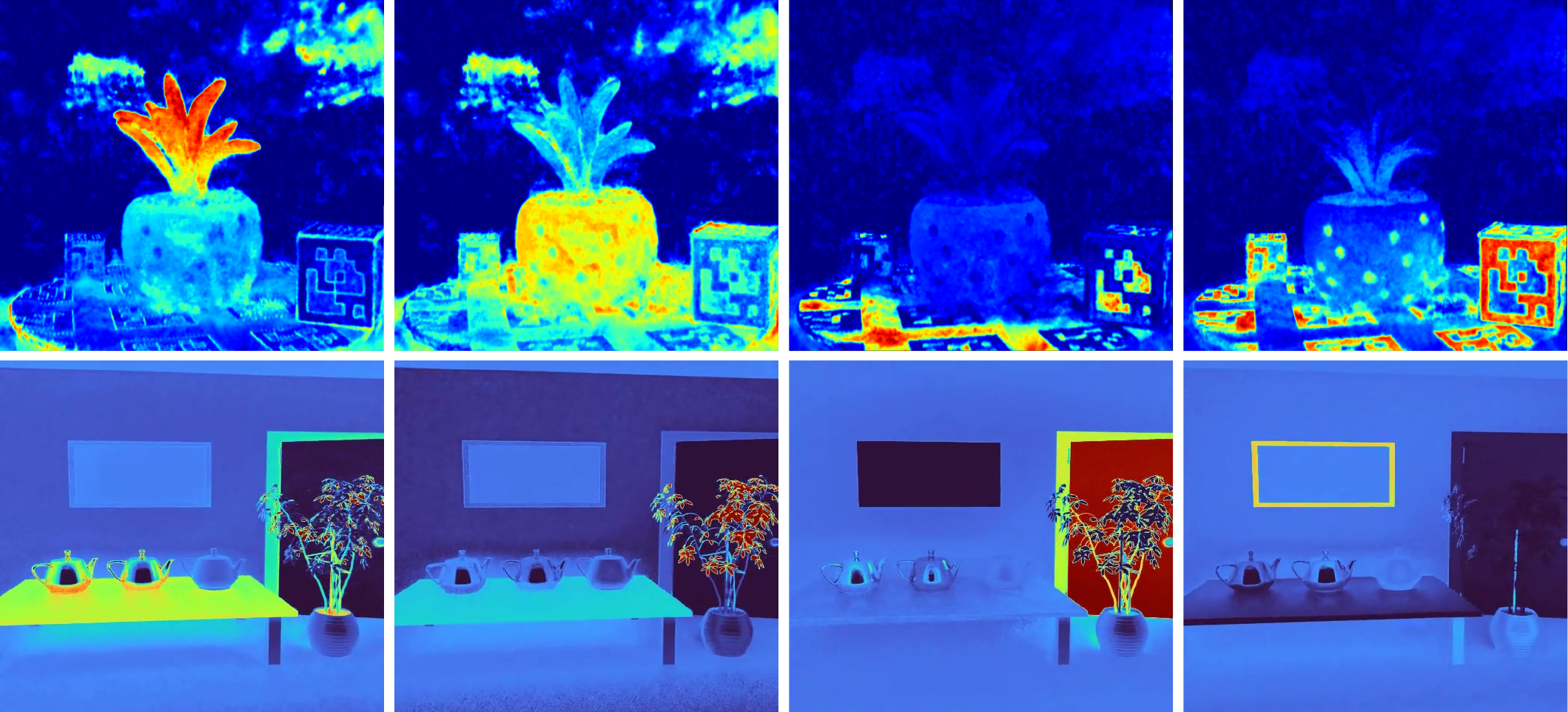}
    \caption{Visualization of the learned material abundance maps for the \textit{Anacampseros} scene from the BaySpec dataset (top row) and the \textit{Ajar} scene from the NeSpoF dataset (bottom row). Each map represents the spatial distribution of a material’s abundance, highlighting distinct materials in the scene.}
    \label{fig:anacampseros_abundances}
\end{figure}

Additionally, Figure~\ref{fig:intro_fig} demonstrates the capabilities of UnMix-NeRF for material segmentation and scene editing. Leveraging the learned endmembers and their corresponding abundance maps, our approach enables precise material separation and targeted modifications to scene appearance. Furthermore, Figure~\ref{fig:anacampseros_abundances} presents a visualization of the learned material abundances for the \textit{Anacampseros} scene in the BaySpec dataset. Each image corresponds to a learned abundance map, highlighting the presence of distinct materials in the scene. The spatial variations observed across the different maps indicate that our approach successfully captures and disentangles material-specific contributions.

%

\subsection{Quantitative Results}

\begin{table}[t]
    \centering
      \resizebox{\columnwidth}{!}{
    \begin{tabular}{lcccc}
        \toprule
        \textbf{Method} & \textbf{Scene} & \textbf{PSNR}~$\uparrow$ & \textbf{RMSE}~$\downarrow$ & \textbf{Time} \\
        \midrule
        HS-NeRF* & avg. & 26.0    & 0.04    & 5 hours \\
        NeSpoF   & avg. & 33.0 & 0.02 & 11.9 hours \\
        \midrule
        \multirow{5}{*}{Ours} 
                 & ajar   & 38.09 & 0.01 & 43 min \\
                 & hotdog   & 34.47 & 0.01 & 45 min  \\
                 & cbox-dragon   & 32.21 & 0.02 & 45 min \\
                 & cbox-sphere   & 27.96 & 0.03 & 44 min \\
                 \cmidrule(lr){2-5}
                 & avg. & \textbf{33.2} & \textbf{0.02} & \textbf{44 min} \\
        \bottomrule
    \end{tabular}
    }
    \caption{Comparison on the NeSpoF dataset. HS-NeRF* and NeSpoF report averaged results (avg.) over 4 scenes; for Ours, per-scene results and the overall averaged metrics are provided.}
    \label{tab:nespof}
\end{table}

\begin{table*}[htp]
  \centering
    Bayspec Dataset
  \resizebox{\textwidth}{!}{
\scriptsize
\begin{tabular}{l cccc cccc cccc} 
  \toprule
  \multirow{2}{*}{Method} & \multicolumn{4}{c}{Pinecone} & \multicolumn{4}{c}{Caladium} & \multicolumn{4}{c}{Anacampseros} \\  
           & \PSNR & \SSIM & \MRAE & \RMSE & \PSNR & \SSIM & \MRAE & \RMSE & \PSNR & \SSIM & \MRAE & \RMSE \\
  \cmidrule(lr){1-1} 
  \cmidrule(lr){2-5} \cmidrule(lr){6-9} \cmidrule(lr){10-13}
  
  NeRF 
  &  22.82
  &  0.6113
  &  0.0446
  &  0.0728
  &  23.12
  &  0.58348
  &  0.0491
  &  0.0709
  &  24.12
  &  0.6220
  &  0.0384
  &  0.0623 \\
  
  MipNeRF      
  & 21.45
  & 0.5738
  & 0.0410
  & 0.0856 
  &  23.36
  &  0.5935
  &  0.0487
  &  0.0685
  &  23.43
  &  0.6160
  &  0.0408
  &  0.0786 \\
  
  TensoRF
  & 24.12
  & 0.6454
  & 0.0593
  & 0.0625
  &  24.79
  &  0.6424
  &  0.0516
  &  0.0577
  &  25.07
  &  0.6569
  &  0.0394
  &  0.0558 \\
  
  Nerfacto
  & 15.36
  & 0.4935
  & 0.0707
  & 0.1709
  &  20.67
  &  0.6208
  &  0.0529
  &  0.0945
  &  21.32
  &  0.6423
  &  0.0417
  &  0.0867 \\
  
  MipNeRF360     
  & 20.93
  & \yellowc0.7355
  & \yellowc0.0279
  & \yellowc0.0507
  &  \yellowc26.93
  &  \yellowc0.7371
  &  \yellowc0.0332
  &  \yellowc0.0461
  &  \yellowc26.73
  &  \yellowc0.7601
  & \yellowc0.0230
  &  \yellowc0.0461 \\
  
  HS-NeRF      
  & 20.07 
  & 0.581 
  & 0.0725 
  & 0.1521
  &  19.084
  &  0.705
  &  0.0533
  &  0.0902
  &  20.32
  &  0.7260
  &  0.0345
  &  0.0789 \\
  
  3DGS 
  & \yellowc22.65 & 0.6039 & 0.0668 & 0.0819 
  & 23.50 &  0.7131 & 0.2889 & 0.0758 
  & 22.59 & 0.5786 & 0.0447 & 0.0853 \\
  
  HyperGS   
  & \orangec27.0 
  & \orangec0.7509 
  & \orangec0.0309 
  & \orangec0.0447 
  & \orangec27.70 
  & \orangec0.8354
  & \orangec0.0271 
  & \orangec0.0414
  & \orangec26.62 
  & \orangec0.7545
  & \orangec0.0183 
  & \orangec0.0460 \\
  
  Ours   
  & \redc27.13 & \redc0.8174 & \redc0.0287 & \redc0.0429
  & \redc30.08 & \redc0.8541 & \redc0.0237 &  \redc0.0312
  & \redc28.20 & \redc0.7612 & \redc0.0154 & \redc0.0392 \\
  
  \bottomrule
\end{tabular}
  }
  \caption{Quantitative results on the BaySpec dataset for the \textit{Pinecone}, \textit{Caladium}, and \textit{Anacampseros} scenes. Our method achieves the best performance across all metrics.}
  \label{tab:full_SOP}
\end{table*}


\begin{table}[htp]
  \centering
  \scriptsize
  \setlength{\tabcolsep}{3pt}
        \resizebox{\columnwidth}{!}{
  \begin{tabular}{l cccc cccc} 
    \toprule
    \multirow{2}{*}{Method} & \multicolumn{4}{c}{Rosemary} & \multicolumn{4}{c}{Basil} \\
           & \PSNR & \SSIM & \MRAE & \RMSE & \PSNR & \SSIM & \MRAE & \RMSE \\
    \cmidrule(lr){1-1} 
    \cmidrule(lr){2-5} \cmidrule(lr){6-9}
    
    NeRF       
    & 8.42 & 0.7461 & 0.0284 & 0.3560 
    & 9.91 & 0.5534 & 0.0769 & 0.5256 \\
    
    MipNeRF      
    & 13.64* & 0.5684* & 1000* & 0.2083* 
    & 10.11 & 0.5878 & 0.0728 & 0.5334 \\
    
    TensoRF   
    & 12.1 & 0.73351 & 0.0212 & 0.2662 
    & 15.23 & 0.5811 & 0.0435 & 0.3628 \\
    
    Nerfacto    
    & 18.66 & 0.8836 & 0.0078 & 0.1205
    & 16.54 & 0.7915 & 0.0176 & 0.1655 \\
    
    MipNerf360   
    & 8.47 & 0.7518 & 0.0876 & 0.3825
    & 13.92 & 0.8584 & 0.0497 & 0.2035 \\
    
    HS-NeRF     
    & *18.60 & *0.887 & *0.0077 & *0.1187 
    & *16.81 & *0.771 & *0.0172 & *0.1587 \\
    
    3DGS   
    & \yellowc25.56 & \orangec0.9695 & \yellowc0.0028 & \yellowc0.0534
    & \yellowc21.19 & \yellowc0.9385 & \yellowc0.0101 & \yellowc0.0897 \\
    
    HyperGS    
    & \orangec26.77 & \redc0.9845 & \orangec0.0021 & \orangec0.0445
    & \orangec25.30 & \orangec0.9503 & \orangec0.00514 & \orangec0.0569 \\
    
    Ours    
    & \redc28.91 & \yellowc0.9355 & \redc0.0019 & \redc0.0332
    & \redc29.21 & \redc0.9584 & \redc0.0043 & \redc0.0364 \\
    
    \bottomrule
  \end{tabular}
  }
  \caption{Quantitative results on the Surface Optics dataset for the \textit{Rosemary} and \textit{Basil} scenes. Our method achieves the best overall performance across all metrics. Results for the \textit{Tools} and \textit{Origami} scenes are provided in the supplementary material.}
  \label{tab:full_SOP_onecol}
\end{table}

Quantitative evaluations of our UnMix-NeRF method on three different datasets demonstrate superior performance across a variety of scenes and conditions. In all tables, the best performing result for each metric is highlighted in \textit{red}, the second-best in \textit{orange}, and the third-best in \textit{yellow}.

Table \ref{tab:nespof} presents the quantitative comparison on the NeSpoF dataset, where our method achieves state-of-the-art performance while maintaining significantly lower computational cost. Our approach reaches an average PSNR of $33.2$ while reducing the computation time to only $44$ minutes. Additionally, we match the RMSE of NeSpoF, which requires over $11.9$ hours per scene. The per-scene results highlight the robustness of our method.

We further evaluate UnMix-NeRF on the BaySpec dataset, as shown in Table \ref{tab:full_SOP}. Our method consistently outperforms all baselines in all scenes, achieving the highest PSNR, SSIM, and the lowest SAM and RMSE values. Notably, our model improves over HyperGS by $+3.15$ PSNR on the \textit{Caladium} scene. Similar trends are observed in the \textit{Pinecone} and \textit{Anacampseros} scenes, where our approach provides both higher reconstruction fidelity and better spectral consistency. These results show that explicitly modeling spectral unmixing in NeRF improves spectral prediction accuracy over per-pixel regression.


Table \ref{tab:full_SOP_onecol} reports the results on the Surface Optics dataset. Our method achieves the best performance in both the \textit{Rosemary} and \textit{Basil} scenes, obtaining a $+2.14$ and $+3.91$ PSNR improvement over HyperGS, respectively. Furthermore, our approach achieves the lowest SAM and RMSE values, confirming its ability to accurately reconstruct spectral details across diverse material compositions. The results for the remaining scenes are provided in the supplementary material.

\begin{figure}[t]
    \centering
    \includegraphics[width=\columnwidth]{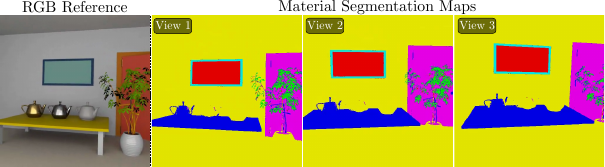}
    \caption{Unsupervised material segmentation of the \textit{Ajar} scene.}
    \label{fig:segmentation}
\end{figure}

Additionally, we evaluate the effectiveness of our method for unsupervised material segmentation on the extended NeSpoF synthetic dataset. Our approach achieves an average F1 score of $0.41$ and a mIoU of $0.28$ across all scenes in the dataset, demonstrating UnMix-NeRF's ability to extract meaningful material clusters without requiring supervision. Figure~\ref{fig:segmentation} shows the unsupervised material segmentation results of our method on the \textit{Ajar} scene.


\section{Conclusions}
\label{sec:conclusions}
We introduced UnMix-NeRF, a novel approach integrating spectral unmixing into neural radiance fields to achieve simultaneous hyperspectral novel view synthesis and unsupervised material segmentation. By modeling spectral reflectance through learned global endmembers and per-point abundances, our method effectively captures intrinsic material properties. Our extensive evaluations show that UnMix-NeRF outperforms existing methods in spectral reconstruction quality and unsupervised material segmentation accuracy. Additionally, our framework supports intuitive material-based scene editing through direct manipulation of endmember dictionaries. 

\vspace{0.1in}
\noindent\textbf{Acknowledgments.} The research reported in this publication was supported by funding from King Abdullah University of Science and Technology (KAUST) - Center of Excellence for Generative AI, under award number 5940.

{

    \small
    \bibliographystyle{ieeenat_fullname}
    \bibliography{main}
}

\end{document}